\documentclass{PoS}
\pdfoutput=1
\usepackage[UKenglish]{babel}
\usepackage{natbib}
\newcommand{\mev}{\, {\rm MeV}}
\newcommand{\be}{\begin{equation}}
\newcommand{\ee}{\end{equation}}
\newcommand{\bea}{\begin{eqnarray}}
\newcommand{\eea}{\end{eqnarray}}

\title{$K \to \pi$ vector form factor with $N_f=2+1+1$ Twisted Mass fermions}

\ShortTitle{$K \to \pi$ vector form factor with $N_f=2+1+1$ Twisted Mass fermions}

\author{ N. Carrasco$^{(a)}$, P. Lami$^{(a,b)}$, V. Lubicz$^{(a,b)}$, E. Picca$^{(a,b)}$,

 \speaker{L. Riggio}$^{(a)}$, S. Simula$^{(a)}$, C. Tarantino$^{(a,b)}$
\\

\it $^{(a)}$ INFN, Sezione di Roma Tre, Rome, Italy. Email: \email{carrasco@fis.uniroma3.it}, \email{lorenzo.riggio@gmail.com}, \email{simula@roma3.infn.it}

\it $^{(b)}$ Dipartimento di Matematica e Fisica, Universit\`a  Roma Tre, Rome, Italy.  Email: \email{lamipaolo@gmail.com}, \email{lubicz@fis.uniroma3.it}, \email{e.picca88@gmail.com}, \email{tarantino@fis.uniroma3.it}

\\

\bf{For the ETM Collaboration}
}

\abstract{ We present a lattice QCD determination of the vector form factor of the kaon semileptonic decay $K\to \pi \ell \nu$ which is relevant for the
extraction of the CKM matrix element $|V_{us}|$ from experimental data. Our result is based on the gauge configurations produced by the
European Twisted Mass Collaboration with $N_f=2+1+1$ dynamical fermions. We simulated at three different values of the lattice
spacing and with pion masses as small as $210$ MeV. Our preliminary estimate for the vector form factor at zero momentum transfer is $f_+(0)=0.9683(65)$, where the uncertainty is both statistical and systematic. By combining our result with the experimental value of $f_+(0)|V_{us}|$ we obtain $|V_{us}|=0.2234(16)$, which satisfies the unitarity constraint of the Standard Model at the permille level.}

\FullConference{The 32nd International Symposium on Lattice Field Theory,\\
		23-28 June, 2014\\
		Columbia University New York, NY}

\begin{document}

%%%%%%%%%%%%%%%%%%%%%%%%%%%%%%%
\section{Introduction and simulation details}
\label{sec:intro}
%%%%%%%%%%%%%%%%%%%%%%%%%%%%%%%

In the Standard Model (SM) the relative strenght of the flavor-changing weak currents is parametrized by the Cabibbo-Kobayashi-Maskawa (CKM) matrix elements. An accurate determination of the CKM matrix elements is therefore crucial both for testing the SM and for searching new physics (NP). 

In this letter we present the determination of the matrix element $|V_{us}|$ from the study of semileptonic kaon (Kl3) decays on the lattice.
This determination is obtained combining lattice results for the $K\to \pi \ell \nu$ form factor $f_+(0)$ with the experimental measure of $f_+(0)|V_{us}|$ extracted from the decay rate of the process. Another possible approach for the determination of $|Vus|$ consists in combining the experimental measurement of pion and kaon leptonic decays with the lattice results for the ratio of decay constants $f_K / f_\pi$. This calculation has been also performed by our collaboration and the results are presented in \cite{DECAYCONSTANTS}.

In this contribution we used the ensembles of gauge configurations produced by the European Twisted Mass (ETM) Collaboration with four flavors of dynamical quarks ($N_f = 2+1+1$), which include in the sea, besides two light mass degenerate quarks, also the strange and the charm quarks. 
The simulations were carried out at three different values of the lattice spacing $a$ to allow a controlled extrapolation to the continuum limit, the smallest being approximately $0.06fm$, and at different lattice volumes. The simulated pion masses used in this analysis range from $210 \mev$ to approximately $450 \mev$.
For each ensemble we used a number of gauge configurations corresponding to a separation of 20 trajectories to avoid autocorrelations.  
The gauge fields were simulated using the Iwasaki gluon action \cite{Iwasaki:1985we}, while sea quarks were implemented with the Wilson Twisted Mass Action \cite{Frezzotti:2003xj}, which at maximal twist allows for an automatic ${\cal{O}}(a)$-improvement \cite{Frezzotti:2003ni}. 
To avoid mixing in the strange and charm sectors we adopted a non-unitary setup in which valence quarks are simulated for each flavor using the Osterwalder-Seiler action \cite{Osterwalder:1977pc}. 
At each lattice spacing different values of light and strange quark masses have been considered to study the dependence of the form factor $f_+(0)$ on $m_\ell$ and to allow for a small interpolation in $m_s$. In our final result for the form factor $f_+(0)$ we used for the physical values of $m_\ell$ and $m_s$ the values obtained in \cite{Carrasco:2014cwa}. 
Valence quarks were simulated at different values of the spatial momenta using Twisted Boundary conditions \cite{Bedaque:2004kc,deDivitiis:2004kq,Guadagnoli:2005be}, allowing us to cover both the spacelike and the timelike region of the $4-$momentum transfer.
For more details on the simulation the reader can consult \cite{Carrasco:2014cwa}.

In the present work we studied a combination of three-points correlation functions in order to extract the form factors $f_+$ and $f_0$ as functions of the squared $4-$momentum transfer $q^2$, the light quark mass $m_\ell$ and the lattice spacing $a$.
The small interpolation in the strange quark mass, which has been simulated at three different values close to the physical one, is addressed with a simple quadratic spline.

Our result is $f_+(0)=0.9683(65)$ where the uncertainty is both statistical and systematic. This allows us to extract the value of the CKM matrix element $|V_{us}|=0.2234(16)$, which is compatible with the unitarity constraint of the Standard Model.

\section{Extraction of the form factors at $q^2=0$}
The matrix element of the vector current between two pseudoscalar mesons decomposes into two form factors, $f_+$ and $f_-$,
\begin{equation}
\label{eq:matrixelementdecomposition}
 \left<  \pi(p')|V_{\mu}|K(p)  \right>=(p_\mu+p'_\mu)f_+(q^2)+(p_\mu-p'_\mu)f_{-}(q^2),
\end{equation}
which depend on the square of the $4-$momentum transfer $q_\mu=p_\mu-p'_\mu$.
The scalar form factor $f_0$ is defined as
\begin{equation}
\label{eq:f0def}
f_0(q^2)=f_+(q^2)+\frac{q^2}{M_K^2-M_{\pi}^2}f_{-}(q^2),
\end{equation}
and therefore satisfies the relation $f_+(0)=f_0(0)$.
The matrix element in Eq.~(\ref{eq:matrixelementdecomposition}) can be derived from the time dependence of a convenient combination of Euclidean three-point correlation functions in lattice QCD. 
As it is well known at large time distances the three-point functions can be written as
\begin{equation}
\label{eq:3pt}
C_\mu  ^{K\pi } \left( {t_x ,t_y ,\vec p,\vec p^{~'}} \right)_{ ~ \overrightarrow{
 t_x  \gg a ~~
 \left( {t_y  - t_x } \right) \gg a} ~ } Z_V \frac{{\sqrt {Z_K Z_\pi  } }} {{4E_K E_\pi  }}\left\langle {\pi \left( {p'} \right)} \right|\; V_\mu  \;\left| {K\left( p \right)} \right\rangle e^{ - E_K t_x  - E_\pi  \left( {t_y  - t_x } \right)} ,
\end{equation}
and therefore they can be combined in the ratio $R_\mu$ 
\begin{eqnarray}
& R_\mu &(t,\vec{p},\vec{p'}) =\frac{ C_{\mu}^{K\pi}(t,\frac{T}{2},\vec{p},\vec{p'} ) C_{\mu}^{\pi K}(t,\frac{T}{2},\vec{p'},\vec{p})}{C_{\mu}^{\pi\pi}(t,\frac{T}{2},\vec{p'},\vec{p'})C_{\mu}^{KK}(t,\frac{T}{2},\vec{p},\vec{p})},  \\
&R_\mu &_{ ~ \overrightarrow{
 t \gg a } ~ } \frac{\left<  \pi(p')|V_{\mu}|K(p) \right> \left< K(p) |V_{\mu}| \pi(p') \right>}{\left<  \pi(p')|V_{\mu}| \pi(p') \right>\left< K(p) |V_{\mu}|K(p) \right>},
\end{eqnarray}
which is independent of the vector renormalization constant $Z_V$ and on the matrix elements $Z_\pi = | \langle \pi | \overline{u} \gamma_5 d | 0 \rangle|^2$ and $Z_K = | \langle \pi | \overline{s} \gamma_5 u | 0 \rangle|^2$.
So the matrix elements $\left<V_0\right>$ and $\left<V_i\right>$ can be extracted from the $R_\mu(t,\vec{p},\vec{p'})$ plateaux as
\begin{eqnarray}
\left<  \pi(p')|V_{0}|K(p)  \right> & = &\left< V_0 \right>=2\sqrt{R_0}\sqrt{EE'},\\
\left<  \pi(p')|V_{i}|K(p)  \right> & = &\left< V_i \right>=2\sqrt{R_i}\sqrt{pp'},
\end{eqnarray}
and used to extract the form factors trough the relations
\begin{eqnarray}
\label{eq:f+&f-}
f_{+}(q^{2})& = &\frac{(E-E')\left<V_i\right>-(p_i-p'_i)\left<V_0\right>}{2Ep'_i-2E'p_i},\nonumber \\
f_{-}(q^{2})& = &\frac{(p_i+p'_i)\left<V_0\right>-(E+E')\left<V_i\right>}{2Ep'_i-2E'p_i}.
\end{eqnarray}
The energies appearing in Eqs.~(\ref{eq:f+&f-}) are extracted from the dispersion relation with the masses obtained fitting the two-points correlation function of pseudoscalar mesons at rest.
Finally $f_0(q^2)$ can be calculated from Eq.~(\ref{eq:f0def}).
\begin{figure}
\begin{center}
\scalebox{0.3}{
\includegraphics{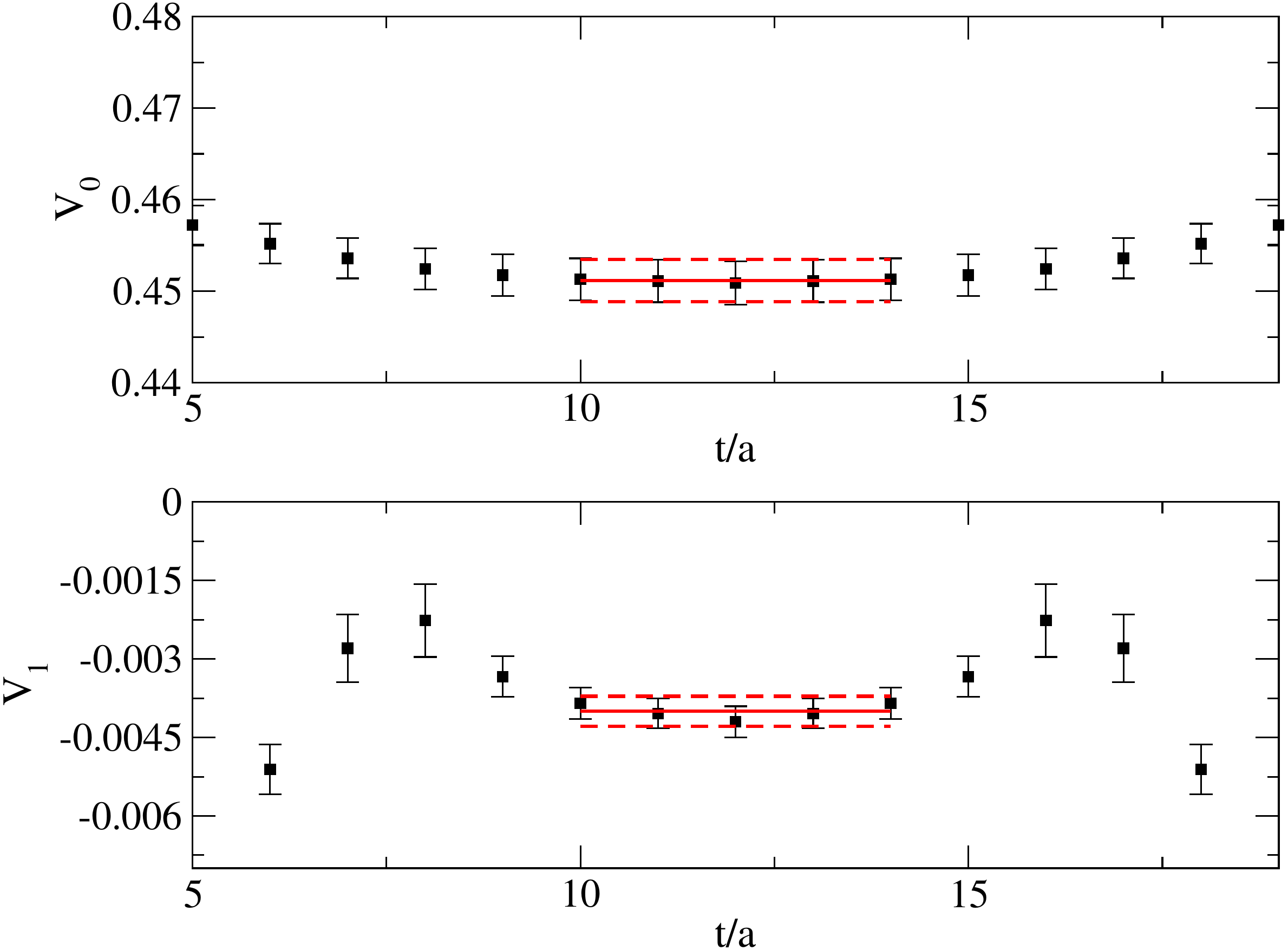}
}
\end{center}
\vspace*{-0.8cm}
\caption{Example of the matrix elements $\left< V_0 \right>$ and $\left< V_i \right>$ extracted from the quantity $R_{\mu}$ corresponding to an ensemble with $\beta=1.90$, $L/a=24$, $a\mu_l=0.0080$, $a\mu_s=0.0225$, $|\vec{p}|=|\vec{p'}|\simeq87$MeV. }
\label{fig:matel}
\end{figure}
An example of the extraction of the matrix elements can be seen in Fig.~\ref{fig:matel}.

The next step was the study of the form factors $f_+$ and $f_0$ as a function of the $4-$momentum transfer to interpolate the data at $q^2=0$. This was done by fitting simultaneously $f_+$ and $f_0$ using the $z-$expansion (Eq.~(\ref{eq:zexpansion})) as parametrized in \cite{Bourrely:2008za}, and by imposing the condition $f_+(0)=f_0(0)$. 
\begin{eqnarray}
\label{eq:zexpansion}
f_ +  (q^2 ) = \frac{{a_0  + a_1 \left( {z + \frac{1}{2}z^2 } \right)}}{{1 - \frac{{q^2 }}{{M^2 _V }}}}, \nonumber \\
f_0 (q^2 ) = \frac{{b_0  + b_1 \left( {z + \frac{1}{2}z^2 } \right)}}{{1 - \frac{{q^2 }}{{M^2 _S }}}}.
\end{eqnarray}
In Eq.~(\ref{eq:zexpansion}) $M_S$ and $M_V$ are the scalar and the vector pole mass respectively, and $z$ is defined as 
\begin{equation}
z = \frac{{\sqrt {t_ +   - q^2 }  - \sqrt {t_ +   - t_0 } }}{{\sqrt {t_ +   - q^2 }  + \sqrt {t_ +   - t_0 } }}
\end{equation}
where $t_+$ and $t_0$ are
\begin{eqnarray}
t_ +  & = & \left( {M_K  + M_\pi  } \right)^2 \nonumber \\
t_0  & = & \left( {M_K  + M_\pi  } \right)\left( {\sqrt {M_K }  - \sqrt {M_\pi  } } \right)^2 .
\end{eqnarray}

\begin{figure}[htb!]
\centering
\scalebox{0.25}{\includegraphics{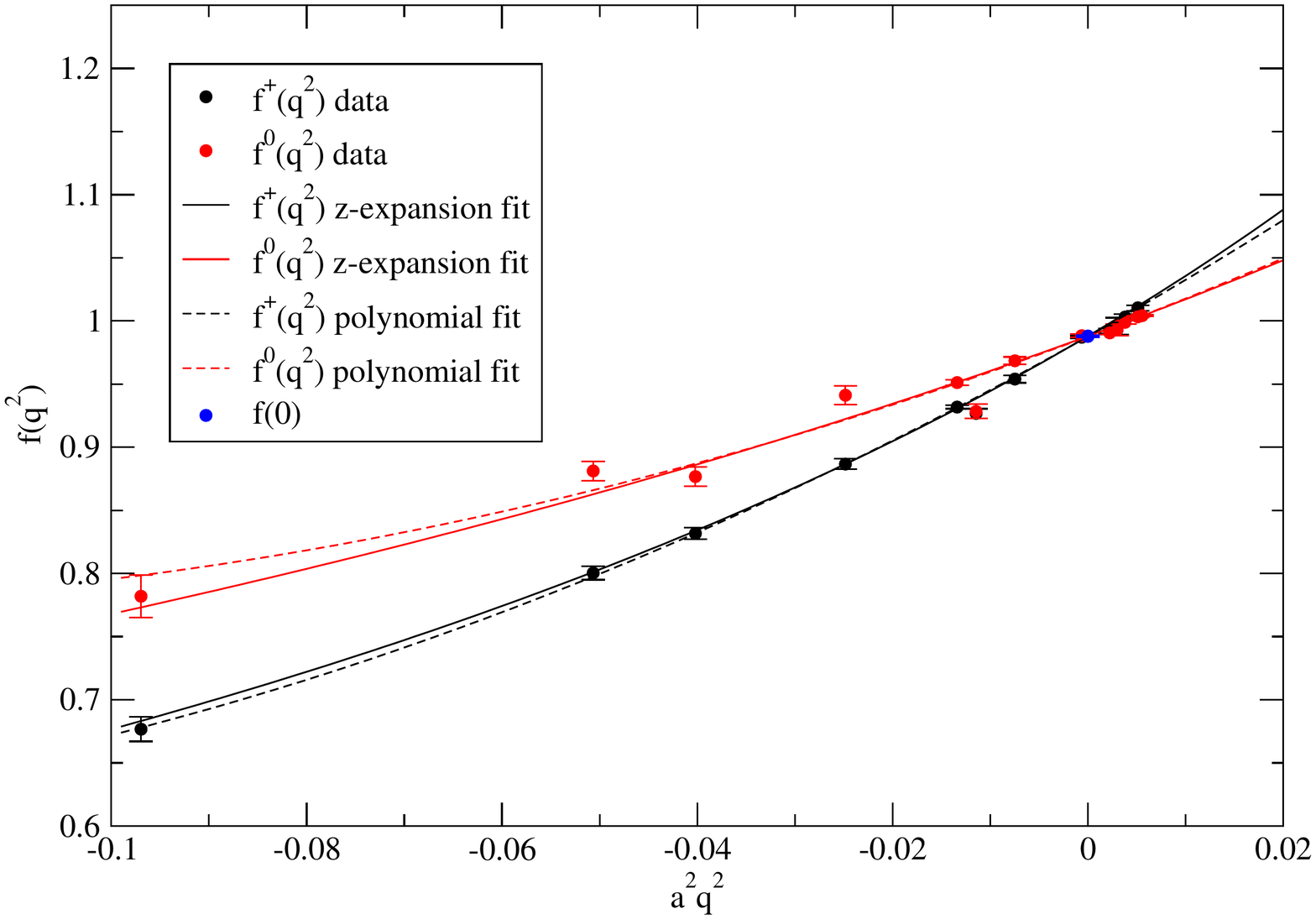}}
\scalebox{0.25}{\includegraphics{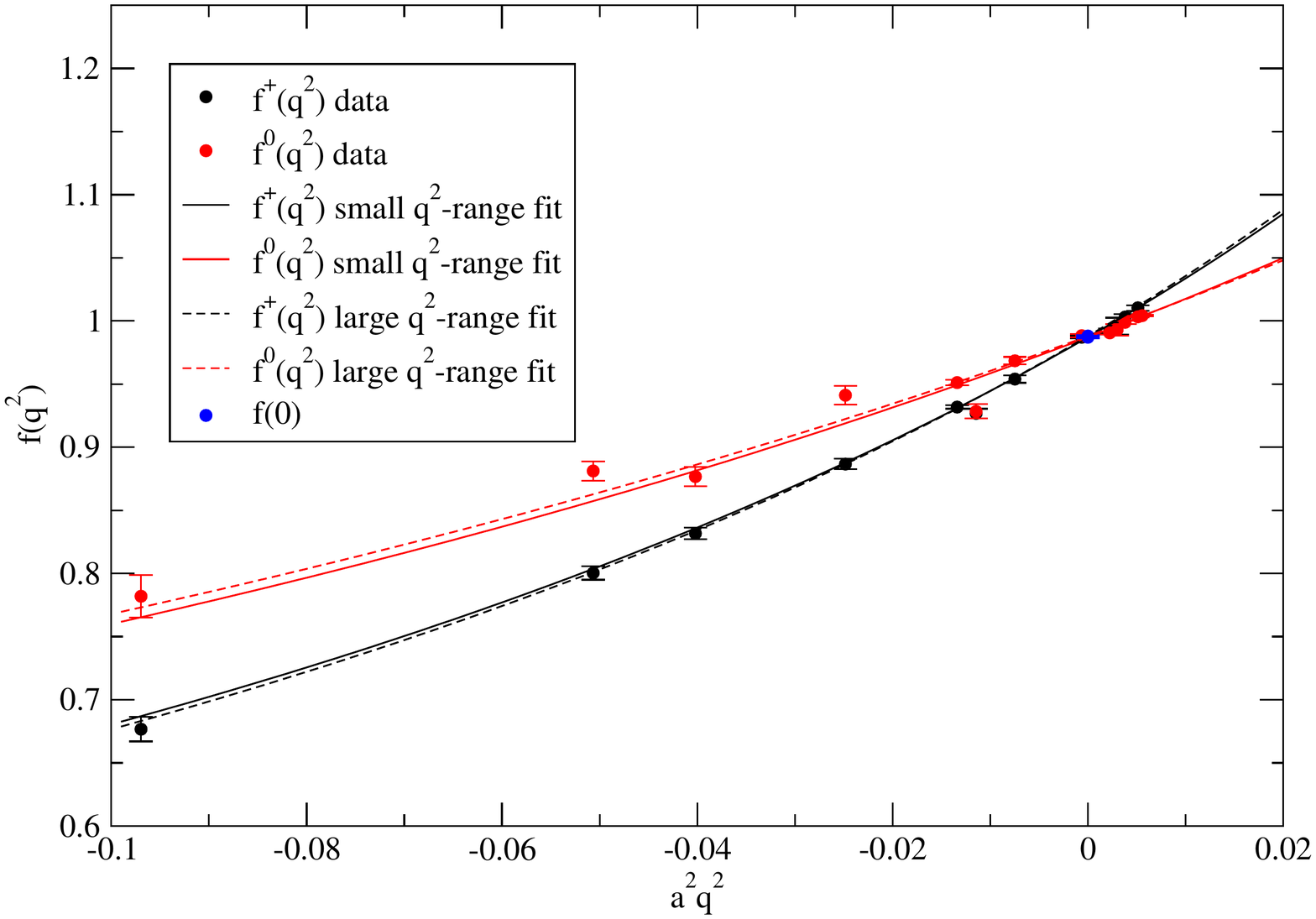}}
\vspace*{-0.8cm}
\caption{\it Left panel: interpolation of the form factors to $q^2=0$ using the $z$ expansion (continuum line) compared to the one obtained with a polynomial fit (dashed line). Right panel: interpolation of the form factors to $q^2=0$ using only the data around $q^2=0$, for this ensemble $a^2q^2 < 0.01$ (continuum line), or a larger range in $q^2$ (dashed line). Both plots corresponds to $\beta=1.90$, $L/a=32$, $a\mu_l=0.0040$, $a\mu_s=0.0225$.
}
\label{fig:q2fit}
\end{figure}

We also tried to fit the $q^2$ dependence using other fit ansatz (e.g. polynomial expression in $q^2$) and including in the fit also the data corresponding to large negative transferred momenta, obtaining nearly identical results as it can be seen in Fig~\ref{fig:q2fit}.

\section{Extrapolation of $f_+(0)$}
In order to compute the physical value of the vector form factor $f_+(0)$, we first performed a small interpolation of our lattice data to the physical value of the strange quark mass $m_s$ determined in \cite{Carrasco:2014cwa}.
Then we analyzed the dependence of $f_+(0)$ as a function of the (renormalized) light-quark mass $m_\ell$ and of the lattice spacing and extrapolated it to the physical point using both an SU(2) and an SU(3) ChPT prediction. Notice however that also the SU(3) fit was performed at fixed physical value of the strange quark mass.\\ 
The SU(2) ChPT prediction at the next-to-leading order (NLO) for $f_+(0)$ \cite{Flynn:2008tg}, reads as follows:
\begin{equation}
\label{eq:SU2fit}
f_ +  (0) = F^ +  _0 \left( {1 - \frac{3}{4}\xi \log \xi  + P_2 \xi  + P_3 a^2 } \right)
\end{equation}
where $\xi_\ell = 2B m_\ell / 16\pi^2f^2$ with $B$ and $f$ being the SU(2) low-energy constants (LECs) entering the LO chiral Lagrangian determined in \cite{Carrasco:2014cwa}. $F^+_0$, $P_2$ and $P_3$, on the other hand, are left as free fit parameters.

In SU(3) ChPT the expression for the vector form factor $f_+(0)$ is the following: 
\begin{equation}
\label{eq:SU3fit}
 f_ +  (0) = 1 + f_2  + \Delta f,  
\end{equation}
where $f_2$ can be written in full QCD \cite{Gasser:1984ux,Gasser:1984gg} as:
\begin{equation}
 f^{{\rm{full QCD}}} _2 = \frac{3}{2}H_{\pi K}  + \frac{3}{2}H_{\eta K},  
\end{equation}
with
\begin{equation}
 H_{PQ}  =  - \frac{1}{{64\pi ^2 f^2 _\pi  }}\left[ {M^2 _P  + M^2 _Q  + \frac{{2M^2 _P M^2 _Q }}{{M^2 _P  - M^2 _Q }}\log \frac{{M^2 _Q }}{{M^2 _P }}} \right]. 
\end{equation}
The quantity $\Delta f$ represents next-to-next-to-leading order (NNLO) contributions and beyond, which in our fit is parametrised as: 
\begin{equation}
\Delta f = \left( {m_s  - m _\ell  } \right)^2\left[ {\Delta _0  + \Delta _1 m_\ell} \right] + \Delta _2 a ^2,
\end{equation}
so that Eq.~(\ref{eq:SU3fit}) verifies the Ademollo Gatto theorem \cite{Ademollo:1964sr} in the continuum limit i.e. deviations from unity are proportional to $(m_s-m_\ell)^2$.

\begin{figure}[htb!]
\centering
\scalebox{0.25}{\includegraphics{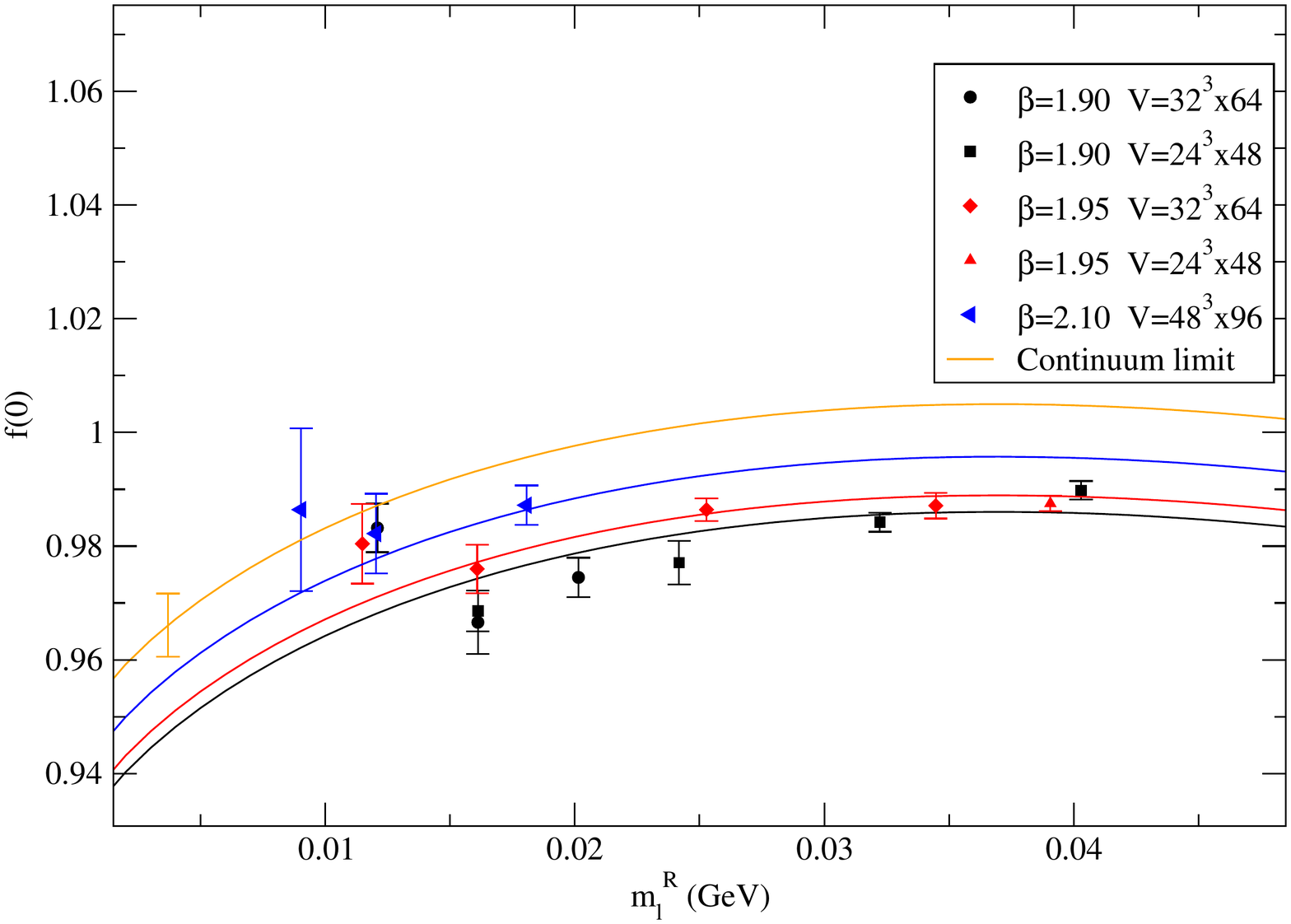}}
\scalebox{0.25}{\includegraphics{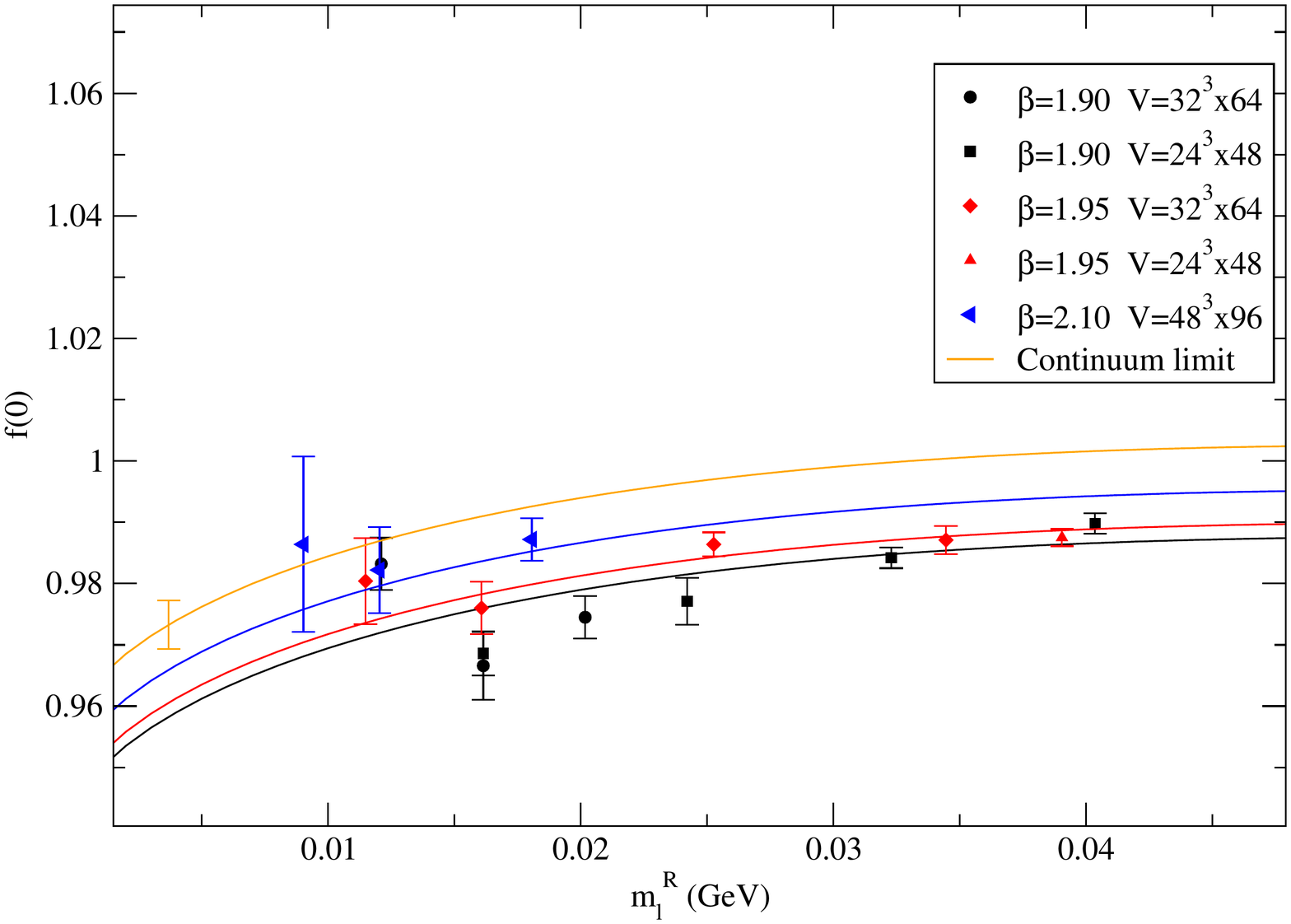}}
\vspace*{-0.8cm}
\caption{\it Chiral and continuum extrapolation of $f_+(0)$ based on the NLO SU(2) ChPT fit given in Eq.~(\ref{eq:SU2fit}) (left) and on the NNLO SU(3) ChPT fit given in Eq.~(\ref{eq:SU3fit}) (right).}
\label{fig:SU2SU3fit}
\end{figure}

The chiral and continuum extrapolations of $f_+(0)$ are shown in Fig.~\ref{fig:SU2SU3fit} for both the SU(2) and the SU(3) fit. 
Combining the two analysis we get our final result for the vector form factor $f_+(0)$:
\begin{equation}
\label{eq:fp0res}
f_+(0)=0.9683(50)_{stat+fit}(42)_{Chir}=0.9683(65),
\end{equation}
where $()_{stat+fit}$ indicates the statistical uncertainty which includes the one induced by the fitting procedure and the uncertainties in the determination of all the input parameters needed for the analysis, namely the values of the light quark mass $m_\ell$, the lattice spacing $a$ and the SU(2) ChPT low energy constants $f$ and $B$, which were determined in \cite{Carrasco:2014cwa}.
The systematic uncertainty in the chiral extrapolation, namely $()_{Chir}$, has been evaluated from the difference in the results corresponding to the two chiral extrapolations we performed. It should be noticed also that the two lattice points calculated at the same lattice spacing and light quark mass but different volumes, as it can be seen in Fig.~\ref{fig:SU2SU3fit}, are well compatible within uncertainties, allowing us to state that finite size effects can be neglected in our analysis.
Combining the present result with the experimental value of $|V_{us}|f_+(0)$ from \cite{Antonelli:2010yf} we can estimate $|V_{us}|$ obtaining:
\begin{equation}
|V_{us}|=0.2234(16). 
\end{equation} 
This value can also be compared with the determination of $|V_{us}|$ from the ratio of leptonic PS decay constants $f_{K^+} /f_{\pi^+}$ that we obtained in \cite{DECAYCONSTANTS} wich reads $|V_{us}|=0.2271(29)$
As a phenemenological application we can use our results to test the unitarity of the first row of the CKM matrix taking the value of $|V_{ud}|$ from the $\beta-$decay \cite{Hardy:2009} and ignoring $|V_{ub}|^2$, which is negligible given the present uncertainties, finding:
\begin{eqnarray}
\label{eq:utest}
 |V_{ud}|^2+|V_{us}|^2+|V_{ub}|^2 &=& 0.9991(8) \hspace*{1.2cm} \rm {from } ~~K_{\ell 3} ~~~\rm{[this~ work]}\nonumber \\
 |V_{ud}|^2+|V_{us}|^2+|V_{ub}|^2 &=& 1.0008(14) \hspace*{1cm} \rm {from } ~~K_{\ell 2} ~~~\mbox{\cite{DECAYCONSTANTS}}
\end{eqnarray}
As it can be seen in Eq.~(\ref{eq:utest}) both determinations confirm the first row unitarity at the permille level.  

\section{An outlook on a possible extension }
As a possible extension of our analysis we provided an estimate of the vector form factors $f_+$ and $f_0$ not only at $q^2=0$, but on the entire $q^2$ region accessible to experiments, i.e from $q^2=0$ to $q^2=q^2_{max}=(M_K-M_\pi)^2$. 
\begin{figure}[hbt!]
\begin{center}
\scalebox{0.3}{
\includegraphics{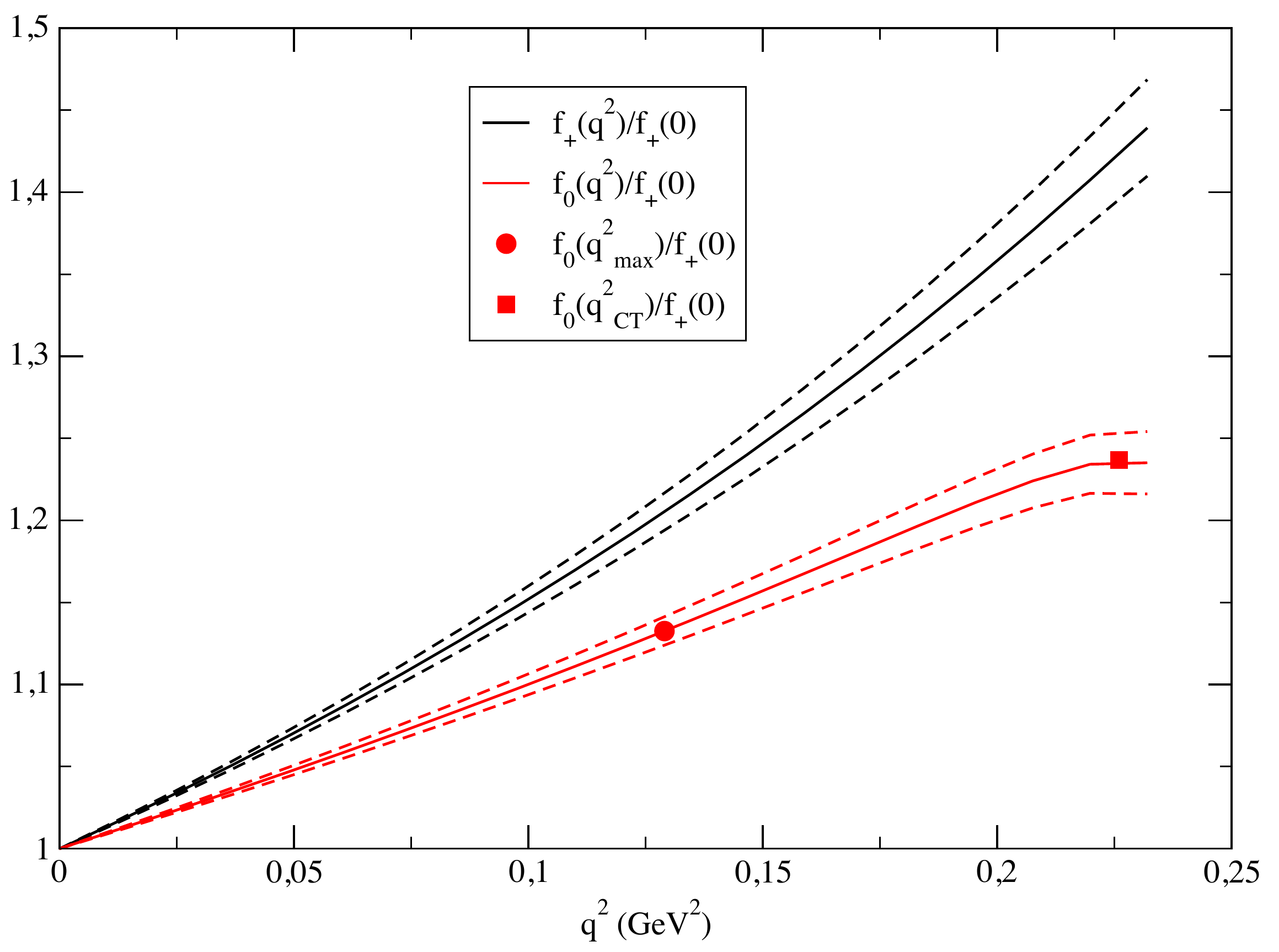}
}
\end{center}
\vspace*{-0.8cm}
\caption{Fit results for the quantities $f_+(q^2)/f_+(0)$ and $f_0(q^2)/f_+(0)$ as functions of $q^2$ at the physical point. The red dot (square) corresponds to $q^2_{max}$  ($q^2_{CT}$). The dashed lines represent the uncertainty of above quantities at one standard deviation.}
\label{fig:f0fpoverfp0}
\end{figure}
To do so we performed a multi-combined fit of the $q^2$, $m_\ell$ and $a$ dependencies of the form factors following the strategy presented in \cite{Lubicz:2010bv}. 
In particular the fit formulas were derived by expanding in powers of $x=M_{\pi}^2/M_K^2$ the NLO SU(3) ChPT predictions for the form factors \cite{Gasser:1984ux,Gasser:1984gg}. Moreover we included in the analysis the constraint from the Callan-Treiman theorem \cite{Callan:1966hu}, which relates the scalar form factor calculated at the unphysical $q^2_{CT}=M_K^2-M_{\pi}^2$ to the ratio of the decay constants $f_K/f_\pi$.
Preliminary result of the form factors in the physical region of $q^2$ are presented in fig. (\ref{fig:f0fpoverfp0}).

%\section{Acknowledgements}

%The authors would like to thank Francesco Sanfilippo for the useful discussions about the work here presented.

\end{document}